%% file: main.tex
\documentclass[letterpaper]{article} 
\usepackage{ijcai26}  
\input{header-ijcai}

\title{Heterogeneity-Aware Client Selection Methodology For Efficient Federated Learning}

\author{
Nihal Balivada$^1$ \and
Shrey Gupta$^2$ \and
Shashank Shreedhar Bhatt$^3$ \and
Suyash Gupta$^1$
\\
\normalfont
$^1$ University of Oregon, US \and $^2$ Boston College, US \and  $^3$ Microsoft Research, India \\
}
\begin{document}

\maketitle

\begin{abstract}
\input{abstract}
\end{abstract}

\input{introduction}

\input{motivation}
\input{background}

\input{related}
\input{overview}
\input{methodology}

\input{experiments}

\input{concl}

\bibliographystyle{named}
\bibliography{ijcai26}

\end{document}

%% file: header-ijcai.tex
\usepackage{times}
\usepackage{soul}
\usepackage{url}
\usepackage[hidelinks]{hyperref}
\usepackage[utf8]{inputenc}
\usepackage[small]{caption}
\usepackage{graphicx}
\usepackage{amsmath}
\usepackage{amsthm}
\usepackage{booktabs}
\usepackage{algorithm}
\usepackage{algpseudocode}
\usepackage[switch]{lineno}
\usepackage{natbib}
\usepackage[table]{xcolor}
\usepackage{enumitem}
\usepackage{multirow}
\usepackage{amssymb}
\usepackage{pifont}

\definecolor{shank}{rgb}{0.7,0.2,0.4}
\definecolor{mygreen}{rgb}{0,0.6,0}
\definecolor{myred}{rgb}{0.8,0,0}

\newcommand{\tf}{Terraform}

%% file: abstract.tex
Federated Learning (FL) enables a distributed client-server architecture where multiple clients collaboratively train a global Machine Learning (ML) model without sharing sensitive local data. 
However, FL often results in lower accuracy than traditional ML algorithms due to statistical heterogeneity across clients.
Prior works attempt to address this by using model updates, such as loss and bias, from client models to select participants that can improve the global model’s accuracy.
However, these updates neither accurately represent a client's heterogeneity nor are their selection methods deterministic.
We mitigate these limitations by introducing \tf{}, a novel client selection methodology that uses gradient updates and a deterministic selection algorithm to select heterogeneous clients for retraining. 
This bi-pronged approach allows \tf{} to achieve up to $47\%$ higher accuracy over prior works.
We further demonstrate its efficiency through comprehensive ablation studies and training time analyses, providing strong justification for the robustness of \tf{}.


%% file: introduction.tex
\section{Introduction}
In this paper, we present {\bf \tf}, a novel client selection methodology for Federated Learning (FL)~\cite{fedavg} that addresses heterogeneity in client data distribution by iteratively partitioning clients into two clusters, easy and hard, based on a split index.
%
An FL framework enables a distributed client-server architecture~\cite{tanenbaum2007distributed} where multiple clients collaboratively train a {\em global} Machine Learning (ML) model without sharing sensitive (local) data.
Each client receives the global model's parameters from the server, uses them to train the model on its data, and shares the updated model parameters with the server.
Consequently, the server aggregates the parameters from various clients. 
This process continues until the global model reaches convergence (highest accuracy).
 
Despite significant seminal research in FL~\cite{yang2019federated}, its practical adoption remains limited due to its lower accuracy yields compared to traditional ML algorithms. 
This is because the local data of clients is often drawn from different distributions, causing \emph{statistical heterogeneity}~\cite{hicsfl}.
In an FL environment, statistical heterogeneity represents the variation in data distributions across clients, caused by the variations in feature or label distributions of clients' local data.
It contributes to the non-iid (non-independent \& identically distributed) nature of the data.

Statistical heterogeneity is a double-edged sword; while it improves model generalization due to data diversity, it increases training complexity and reduces model accuracy.
For example, modeling air quality requires training an ML model on environmental features, such as weather conditions and the topography of the location of interest~\cite{cheng2018neural}. 
Coincidentally, several organizations perform such an air quality modeling on a global scale that spans multiple countries~\cite{galmarini2011aqmeii,pan2017crowdsensing}.
However, globally scaling these models requires data sharing, which is challenging due to proprietary concerns or national security.
FL is an ideal candidate for training such a model, as it enables localized data collection and training.
Unfortunately, FL inadvertently induces statistical heterogeneity, as in this example -- one training location may be an urban area with high pollutant emissions, while another is in a rural setting. 
Although this heterogeneity generalizes the global model, it increases the probability of capturing unwanted patterns, reducing model performance. 

Prior works attempt to mitigate the negative impact of statistical heterogeneity by designing client selection methodologies that decide which clients should participate in FL training. 
\citet{oort} estimate the statistical utility of each client based on multiple factors like model loss, device speed, and bandwidth, and 
then randomly select a subset of high utility clients for retraining.
\citet{poc} select ``$m$'' clients with highest local training loss for retraining, where the value of $m$ is determined by the system.
In contrast, \citet{hicsfl} cluster clients into groups using their final layer's biases and retrain clients in those clusters that have more uniform data distributions.

Although these prior works present valuable designs, they are unable to attain higher accuracy for the following two reasons: 
(i) they assume client updates are either losses or final-layer biases, which do not reveal all details about a client’s data distribution; and 
(ii) their algorithms for selecting clients to be retrained are not deterministic. 
In contrast, our methodology, {\bf \tf}, aims to {\bf eliminate randomness} by deterministically selecting which clients need retraining and 
utilizing {\bf updates that precisely account for statistical heterogeneity}.
Additionally, \tf~ensures that it does {\bf not introduce any new computational or communication costs}.

So {\em how does \tf~meet these goals?}
\tf's novel client selection methodology expects each client to send the {\bf final-layer gradient updates} returned by its model after training and the {\bf dataset size}, which capture statistical utility of a client's data.
\tf~uses gradient updates to {\bf sort the clients} and the dataset size to determine the {\bf Inter-Quartile Range}, which 
enables us to find an optimal {\bf split index} to partition clients into two clusters: {\bf easy} and {\bf hard}.
\tf~{\bf selects and retrains} clients in the hard cluster (iterates over the above steps in the process) 
till only a {\em threshold} number of clients remain.

To ensure that Terraform does perform in practice, we follow prior works and implement it in two popular FL algorithms, FedAvg and FedProx, and
extensively compare Terraform against five state-of-the-art client selection methodologies.
Our results illustrate that Terraform outperforms all the client sampling techniques on the five popular FL datasets (FEMNIST, CIFAR10, CIFAR100, FMNIST, and Tiny ImageNet) and increases accuracy of FedAvg and FedProx by up to $47\%$ and $41\%$, respectively.
Additionally, we validate robustness of Terraform's design through rigorous ablations and training time experiments.

%% file: motivation.tex
\section{The Case for Data Heterogeneity}
\label{s:motivation}
We motivate the need to tackle client data heterogeneity through the following example.
US communities regularly face disproportionate threats of disasters like wildfires and fire incidents~\cite{fema1997rural,psu2022oregondrinkingwater,nifc2024annualreport}.
In response, US government has accelerated installation of IoT wildfire sensors like PTZ Camera~\cite{alertwildfirecamera},
promoted use of smart home sensors like Google Nest~\cite{nestprotect2025}, and funded setup of Hazard workflows like AlertWildfire~\cite{alertwildfire}.
These hazard workflows collect observations with sensors, disseminate those observations to remote cloud for processing, 
infer the impacts of those hazards at cloud, and notify various stakeholders.
However, these hazard workflows face two major challenges:
(i) they collect user data through sensors and upload them to the cloud, which violates user privacy~\cite{fema2013smartphone,raffeg2025legal,nasa2025photos,dhs2019pcr}, and
(ii) they consume massive network bandwidth due to sensor-cloud communication.
This presents an elegant opportunity to leverage FL for disaster response~\cite{jin2024federated};
the sensors can play the role of clients and participate in FL training.

Notice that the two type of sensors we consider here, PTZ camera and Google Nest, have distinct deployment range and alert frequency.
PTZ cameras are sparse, remote (often in forests), and generate few but highly reliable wildfire alerts. 
Nest sensors, on the other hand, are abundant in homes and produce frequent non-wildfire alerts with rare true positives due to smoke from wood stoves, prescribed burns, vehicles, etc.
Since most training data comes from home sensors, any FL system that does not account for data heterogeneity will 
overlook the true positive alerts from the cameras, causing the global model to have low accuracy and missing early wildfire signals.

\tf{} aims to train on these heterogeneous data and ensure that it yields high accuracy for all the events.

%% file: background.tex
\section{Background}
The seminal FedAvg algorithm~\cite{fedavg} coined the term Federated Learning (FL), where in a distributed client-server architecture, multiple clients collaboratively train a global ML model without sharing local data with the server.
We formalize its design as follows:

Consider a standard FL system where a server coordinates $K$ participating clients, indexed by $k = \{1, 2, \ldots, K\}$. 
Each $k$-th client has a local classification dataset (used for training), 
$D_{\text{train}}^k = \{ (x^{(k)}_j,\, y^{(k)}_j) \},$ where $x^{(k)}_j \in \mathbb{R}^m$ is the feature vector, and $y^{(k)}_j \in \{ 1, \ldots, C \}$ are the labels with $C$ classes.
Each client computes a model update by minimizing loss function $L(\theta;\, x^{(k)}_j,\, y^{(k)}_j)$ where \( \theta \) denotes the model parameters. 
These updates are sent to the server, which performs weighted averaging over the received updates, where weights are proportional to each client’s dataset size, \( |D_{\text{train}}^k| \); this leads to updated global model parameters. 
The server then shares the updated model parameters with the clients, and this process continues for either a pre-defined number of rounds or until the model reaches an accuracy saturation.
The accuracy of the global model is evaluated on each client’s test data, \(D_{\text{test}}^k\).

\paragraph{\textbf{Techniques addressing statistical heterogeneity}}
Prior works have explored a variety of approaches to handle statistical heterogeneity in federated learning (FL). \textbf{Regularization-based} techniques discourage client model drift from the global solution by modifying the local objective. For example, FedProx~\cite{fedprox} adds a proximal term for more stable convergence, while FedDyn~\cite{feddyn} aligns the global and client optima via a dynamic regularizer, and FedDC~\cite{feddc} uses an auxiliary variable to decouple and correct gradient and parameter drift by tracking local drift. \textbf{Variance reduction and drift correction} algorithms like SCAFFOLD~\cite{scaffold} remove client drift and achieve comparable convergence rates to centralized SGD using control variates, while FedNova~\cite{fednova} eliminates objective inconsistency due to heterogeneity in the number of local updates by normalizing the client updates. Other algorithms like VRL-SGD~\cite{vrlsgd}, FedVRA~\cite{fedvra} and FedRed~\cite{fedred} reduce variance via gradient tracking or dual variables. \textbf{Adaptive optimization and dynamic aggregation} methods improve performance under statistical heterogeneity by extending techniques from adaptive optimizers like Adagrad, Adam and Yogi to the federated setting~\cite{adaptivefed}. For example, FedAW~\cite{fedaw}, FedAWA~\cite{fedawa} and FedADp~\cite{fedadp} make use of client contributions to adaptively adjust aggregation weights. \textbf{Layer‑wise and architecture‑aware aggregation} methods such as FedMA~\cite{fedma} use neuron-matching and averaging to improve performance on deep CNN and LSTM architectures. \textbf{Contrastive approaches} like MOON~\cite{moon} align global and local representations using contrastive losses.

\textbf{Personalization / multi-mode} techniques tackle statistical heterogeneity via knowledge preserving cross-client transfer and create models tailored for each client. These techniques include using locally regularized models personalization (Ditto~\cite{ditto} with local copies and pFedMe~\cite{pfedme} with Moreau envelopes), meta learning for fast adaptation (PeFLL~\cite{pefll}, Per-FedAvg / MOCHA~\cite{perfedavg} and federated MAML~\cite{fedmaml}), representation decoupling (LG-FedAvg~\cite{lgfedavg}, FedPer~\cite{fedper}, FedRoD~\cite{fedrod} and FedRep~\cite{fedrep}), parameter/feature alignment (FedAS~\cite{fedas} and FedPAC~\cite{fedpac}), bilevel optimization methods (pFedHN~\cite{pfedhn} and FedBabu~\cite{fedbabu}), and partitioning the hypothesis space into compatible submodels (FLOCO~\cite{floco} with solution-simplex methods). \textbf{Clustered and model-mixture} methods cluster clients based on similarity (e.g., FedClust~\cite{fedclust}, CFL~\cite{cfl}, and IFCA~\cite{ifca}) and train a model for each cluster. \textbf{Data augmentation and distillation} techniques like FedDF~\cite{feddf}, FedMD~\cite{fedmd}, FedMix~\cite{fedmix}, FedFed~\cite{fedfed}, FedProto~\cite{fedproto} and FADA~\cite{fada} reduce data statistical heterogeneity by using proxy data or shared statistics.

Finally, \textbf{client selection and sampling} techniques balance efficiency and representativeness by systematically picking the clients participating in each round of federated training. These techniques prioritize clients based on a quantification of  their utility — statistical PoW-D~\cite{poc}, system-based (FedCS~\cite{fedcs}), or both (Oort~\cite{oort}) — towards the expected progress per round. They also consider the diversity or correlation between client updates (DivFL~\cite{divfl} and FedCor~\cite{fedcor}) and population structure captured by clustered sampling~\cite{clustered}, including advanced hierarchical variants like HiCS-FL~\cite{hicsfl}, which additionally uses output-layer bias update information for heterogeneity-aware client selection.

%% file: related.tex
\section{Related Work}

\paragraph{Client Selection Methodologies.}
The FedAvg algorithm selects participating clients randomly in each training round. 
However, this random selection can lead to client drift—a phenomenon in which the global model either converges slowly or becomes trapped in local optima, often due to the repeated selection of highly heterogeneous clients in each round~\cite{scaffold}. 
FedProx~\cite{fedprox} extends FedAvg and addresses client drift by adding a regularization term that penalizes large deviations from the global model. 
This keeps the local updates more aligned with the global optima as well as improves the stability in model training. 
As a result, a broad body of federated learning research focuses on designing principled client selection methodologies that can recognize client heterogeneity and concentrate retraining efforts on the most impactful clients.

\citet{oort} introduce \emph{Oort}, which estimates the statistical utility of each client in FL training based on how effectively its updates improve the global model.
This utility is measured as a combination of model loss, device speed, and bandwidth, and
helps to assign each client a selection probability. 
Oort sorts the clients based on these probabilities and randomly selects a subset of high-probability clients.
Additionally, it randomly selects a few low-probability clients to reduce training bias. 
\citet{poc} introduce \emph{power of choice (PoC)}, which selects clients with high local training loss. 
The server first randomly samples a pool of candidate clients and queries their local training loss. 
Next, it selects $m$ clients with the highest losses, where the value of $m$ is determined by the system.
{\em HiCS-FL}~\cite{hicsfl} uses bias values from the final layer of client models to determine the complexity of its dataset.
It uses these biases to cluster clients into groups, assigns them a sampling probability, and trains clients in clusters with high sampling probability. 
Specifically, HiCS-FL targets clients with less complexity, which is opposite to our goal.
Note: while PoC and Oort use loss as client updates, HiCS-FL uses biases. 
Further, these protocols do not collect updates in each round, which makes them outdated and prevents them from giving a holistic representation of the client's local model.

In contrast, Terraform requires clients to send final-layer {\em gradient updates}, which is the sum of both weights and biases, to compute a client's statistical heterogeneity. 
These gradient updates are a better indicator of the client's distribution due to their proximity to the output layer, and they capture a more direct signal than loss of how well a client's model has learned~\cite{zeiler2014visualizing}.

\paragraph{Hierarchical Splitting.}
The hierarchical selection approach in \tf{} is inspired by the CART (Classification and Regression Trees) model~\cite{CART}, which recursively partitions data at decision nodes into \emph{left} and \emph{right} child nodes using a partitioning criterion.
Hierarchical sampling has also been applied in the meta-learning setting for spatio-temporal domains~\cite{xie2021statistically,taskranking},
where prior works focus on applying hierarchical splitting in a centralized setup. 
In contrast, Terraform introduces hierarchical splitting in a federated (i.e., distributed) learning setting, 
where only gradients are exchanged between clients and the server--minimizing communication overhead and preserving privacy. 
Furthermore, Terraform explicitly accounts for local dataset heterogeneity during client-side training, improving model performance in non-iid scenarios.

%% file: overview.tex
\begin{figure*}[t]
    \centering
    \includegraphics[width=0.95\textwidth]{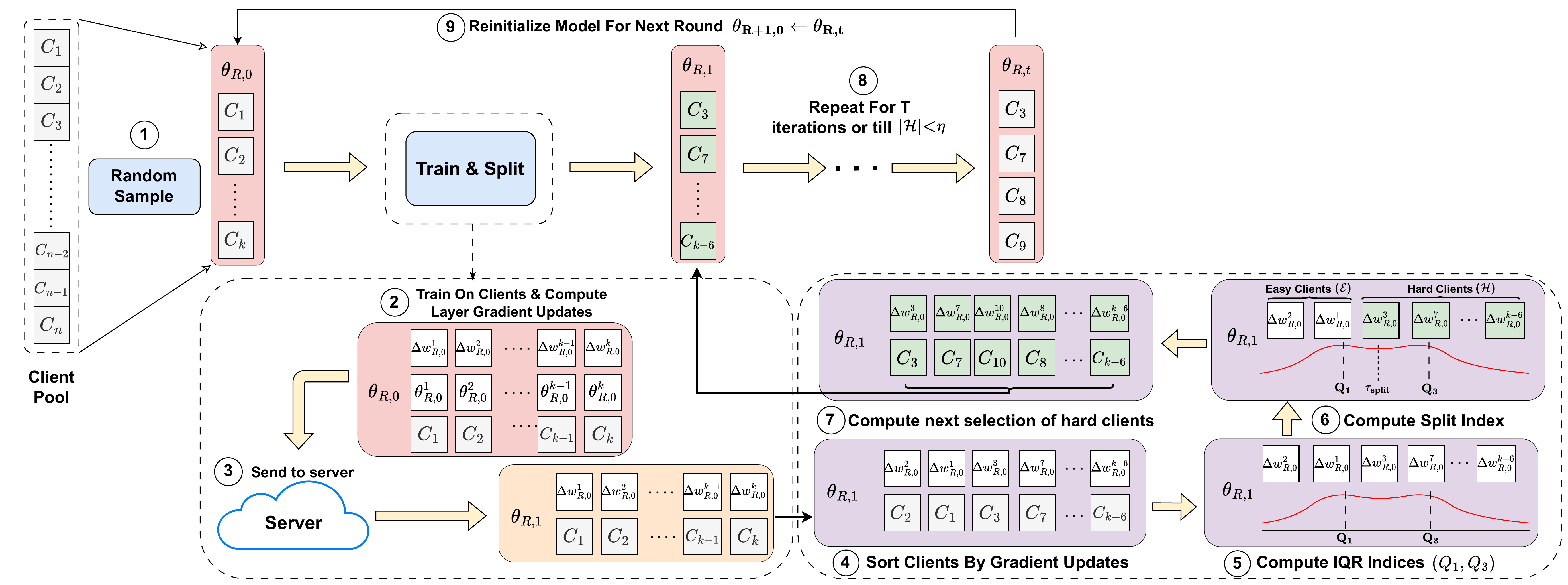}
    \captionsetup{skip=2pt}
    \caption{Terraform Framework.}
    \label{fig:terraform-overview}
\end{figure*}


\section{Terraform Overview}
Terraform is a novel client selection methodology, which when employed by an existing federated learning (FL) algorithm helps it to select statistically heterogeneous clients that negatively impact the accuracy of the FL algorithm.
We use Figure~\ref{fig:terraform-overview} to illustrate the design of Terraform. 

Like any other methodology, Terraform requires rounds to train the clients.
However, each round of Terraform performs hierarchical client selection through a series of iterations.
A single iteration of Terraform includes the following steps. 
The server starts with the global model%
\footnote{Like prior works, we generate the initial global model through random initialization~\cite{fedprox}.}
and sends its parameters to a subset of clients, selected uniformly at random (Step 1 in Figure~\ref{fig:terraform-overview} $\rightarrow$ \ding{172}). 
Next, each client locally trains this model on its dataset \ding{173}, after which it sends the final-layer gradients of the model to the server \ding{174}. 
Upon receiving these gradients from all selected clients, the server aggregates them and begins the process of identifying statistically heterogeneous clients. 
To do so, the server first sorts the clients based on the values of their gradients \ding{175}. 
Next, it uses the dataset size of each client to determine the inter-quartile range (IQR), which is the difference between the third and first quartile, representing 50\% of the gradient magnitudes \ding{176}. 
Within this IQR, the server determines the split index that partitions the clients into easy and hard clusters \ding{177}. 
The server initiates retraining of the clients in the hard cluster (initiating the next iteration) by forwarding them aggregated gradients, and these clients again undergo the above steps \ding{178}.
In essence, this process continues until the number of clients in the hard cluster falls below a predefined threshold \ding{179}.

%% file: methodology.tex
\section{Methodology}
We denote the FL algorithm employing Terraform as $\mathcal{A}$, the initial global model at the server $\mathcal{S}$ as $\theta_{0,0}$, and the number of training rounds as $R$. 
In each round, $r \in \{0,\dots,R-1\}$, $\mathcal{S}$ randomly samples $K$ clients, forming the client pool $\mathcal{C}_{r}^{K}$. 

Within each round, Terraform performs at most $T$ iterations, $T \geq 1$. 
The updated global model parameters at the end of the $t$-th iteration are $\theta_{r,t}$. 
Within each iteration, the clients are split into an easy cluster $\mathcal{E}$ and a hard cluster $\mathcal{H}$ based on a split index, $\tau_{split}$, 
which is determined using the client's gradient update, $\Delta w_{r,t}^{k}$, and its dataset size, $|D_{\text{train}}^k|$; where $k$ represents a client.

The split index, $\tau_{split}$, is selected from the client indices falling between the first ($Q_1$) and third ($Q_3$) quartiles of the distribution of dataset sizes, $|D^k_{train}|$.
Hence, Terraform searches for an optimal split index to generate two groups -- one on which the global model approximates easily (low $|\Delta w_{r,t}^{k}|$, i.e., easy cluster, $\mathcal{E}$) and the other on which the global model struggles to approximate (high $\Delta w_{r,t}^{k}$, i.e., hard cluster, $\mathcal{H}$).
Terraform continues this round until the number of clients in the hard cluster falls below the predefined threshold, $\eta$. 

Next, we explain in detail Terraform's client selection methodology and use Algorithm \ref{alg:terraform} to illustrate its algorithm.

\subsection{Client Model Updates}
In contrast to prior work that relies on losses or final-layer biases for client selection, Terraform leverages final-layer gradients, as these provide deeper insights into the underlying data distribution of each client (refer to Figure~\ref{fig:gradient-ablation}).
As illustrated in Figure~\ref{fig:terraform-overview}, in an iteration $t$ of a round $r$, the client $k$ performs training on its local data $D^{k}_{train}$ using the global model parameters $\theta_{r,t}$, which generates the local model parameters, $\theta_{r,t}^{k}$ such that:
\begin{equation}
\theta_{r,t}^{k} \leftarrow \theta_{r,t} - \Delta w^{k}_{r,t}        
\end{equation}
where $\Delta w^{k}_{r,t}$ is the gradient update between $\theta_{r,t}$ and $\theta_{r,t}^{k}$ due to local training and includes both the weight and bias matrices obtained from the model's final layer (also called the classification layer)~\cite{zeiler2014visualizing}.
The magnitude of the gradient updates $|\Delta w^{k}_{r,t}|$ and the dataset size, $|D_{train}^k|$, are returned to the server. 
The magnitude of a client's total gradient update, $|\Delta w^{k}_{r,t}|$, is obtained by taking the square root of the sum of the squared Frobenius norms~\cite{frobenius} of its final layer's individual parameters' gradient update, $\Delta p^k_i$:
\begin{equation}
|\Delta w_k| = \sqrt{\sum_{i=1}^{L} \|\Delta p^k_i\|_F^2} 
\end{equation}
where L is the number of trainable parameters in the final layer.
The Frobenius norm of a matrix $\mathbf{A} \in \mathbb{R}^{m \times n}$ tells us the size of $\mathbf{A}$, and is given by the square root of the summed absolute squares of its elements:
\begin{equation}
    \|\mathbf{A}\|_F = \sqrt{\sum_{i=1}^{m}\sum_{j=1}^{n} |a_{ij}|^2} 
\end{equation}
Intuitively, the gradient updates indicate how much new information the client’s data is adding to the global model. 
A small gradient update suggests that the global model requires less fine-tuning on the client's data, that is, the client's data distribution aligns with the data distribution learned by the global model.

\subsection{Hierarchical Selection}
Terraform's hierarchical selection takes inspiration from the Classification and Regression Trees (CART) algorithm, which constructs a binary decision tree by recursively partitioning data at each node into \emph{left} and \emph{right} child nodes based on a partitioning criterion.
For classification tasks, well-known partitioning criteria like Gini index and Entropy~\cite{hastie2001elements} are employed, whereas for regression tasks, the Sum of Squared Errors (SSE) method is used.
For example, assume a regression problem that needs to generate a CART decision tree. 
The partition at each node $n_i$ of the tree, where $n_i \in \{1,\ldots, N\}$ is determined using the total SSE of the resulting child nodes defined as $\text{SSE}(N_L)$ and $\text{SSE}(N_R)$.
It equates to, $\text{SSE}(N_{L}) = \sum_{n_i \in N_{L}} (n_i - \bar{n}_{L})^2$, where \( \bar{n}_L \) represents the mean of the left partition; and similarly for the right partition. 
CART selects the partition that minimizes the total SSE for each node, $\text{SSE}_{\text{total}} = \text{SSE}(N_L) + \text{SSE}(N_R)$.

Terraform's hierarchical selection extends these rules to the FL environment such that a client's gradient update magnitude $|\Delta w^k_{r,t}|$ represents a node. 
These gradients are used to calculate the split index that partitions them into two clusters: easy $\mathcal{E}$ and hard $\mathcal{H}$.
However, finding the optimal split index from this large set of gradient magnitudes is hard. 
Thus, Terraform introduces the notion of Inter-Quartile Range (IQR), explained next.

\subsubsection{Inter-Quartile Range (IQR).}
\label{sec:iqr}
Prior to calculating the split index, Terraform sorts the clients based on the magnitude of their gradient updates $|\Delta w^k_{r,t}|$.
Once the clients are sorted, their dataset sizes, $|D^k_{train}|$, are used to calculate the running sum,  $S_k = \sum_{j=1}^{k} |D_{\text{train}}^j|$, where $S_k$ denotes the cumulative dataset size up to the $k$-th client in the list.
This running sum is used to identify a subset of clients that can optimize the calculation of the split index, which we denote as the \emph{Inter-Quartile Range} (IQR). 

The IQR is the difference between the third quartile (Q3) and the first quartile (Q1), representing the middle $50\%$ of the data.
It captures the central distribution of the gradient magnitudes, thereby excluding outliers.
To locate the first quartile index $k_{\text{Q1}}$, we find the smallest $k$ such that $S_k \geq 0.25 \cdot S_K$, where \( S_K \) is the total running sum.
Similarly, the third quartile index $k_{\text{Q3}}$ is the smallest $k$ such that $S_k \geq 0.75 \cdot S_K$.
In Figure \ref{fig:iqr-ablation}, we empirically validate the calculation of the split index with and without IQR.

\subsubsection{Split Index.}
To calculate the optimal split index, we consider it as a problem of splitting a list of scalar values $U = \{u_1,...u_i,...,u_N\}$, where $u_i$ corresponds to gradient magnitude, $|\Delta w^i_{r,t}|$; these are sorted in ascending order, into two distinct subsets. 
\citet{xie2021statistically} shows that the optimal split index $\tau_{split}$
can be obtained by minimizing the variance within the two resulting clusters, which we define as {\em intra-split variance}, $\operatorname{Var_{intra}}$.
Based on this, the optimal splitting index, $\tau_{split}$ can be defined as,
\begin{align}
\tau_{split} &=\underset{1 \le \tau < N}{\operatorname{argmin}} \left( \operatorname{Var}_{\text{intra}}(U_1,U_2) \right) 
\end{align}
where $U_1 = \{u_1, \dots, u_{\tau}\}$, $U_2 = \{u_{\tau+1}, \dots, u_N\}$ are the clusters of $U$ after splitting at a given split index $\tau$.
Furthermore, we can define $\operatorname{Var_{intra}}$ as, 
\begin{align}
\operatorname{Var}_{\text{intra}}(U_1, U_2) &= \frac{|U_1|}{N} \cdot \text{Var}(U_{1}) + \frac{|U_2|}{N} \cdot \text{Var}(U_{2})
\end{align}
where $\operatorname{Var}(U)=\frac{1}{S_N} \sum_{i=1}^{N} (|D^i_{train}| \cdot (u_i - \bar{u})^2$) is the variance of the cluster, \( \bar{u} = \frac{1}{S_N} \sum_{i=1}^{N} |D^i_{train}| \cdot u_i \) is the mean of the cluster, $S_N$ is the total running sum and $N$ is the total number of cluster elements.
Hence, the problem is reduced to an iterative search through the possible split indices $\tau$ to minimize $\operatorname{Var_{intra}}$. 
This minimization, combined with IQR, generates $\tau_{split}$.

We can also frame this as a problem of maximizing the variance between the two clusters, defined as {\em inter-split variance}, $\operatorname{Var}_{inter}$, due to an inverse relation between $\operatorname{Var}_{inter}$ and $\operatorname{Var}_{intra}$; the Law of Total Variance~\cite{fisher1958statistical}.
\begin{align*}
\operatorname{Var}(U) &= \operatorname{Var_{inter}(U_1, U_2)} + \operatorname{Var_{intra}(U_1, U_2)}
\end{align*}
Hence, determining the split index that minimizes $\operatorname{Var_{intra}}$ equates to determining the index that maximally separates clients into two distinct clusters.

This splitting to determine hard clusters is hierarchically repeated until one of these termination conditions is met: (i) reaching the minimum client threshold, $\eta$, or (ii) maximum iteration depth, $T$. 
The model then begins the next round with a randomly sampled set of clients. 
In the following section, we present the complete procedure in Algorithm \ref{alg:terraform}.

\begin{algorithm}[t]
\caption{Terraform Algorithm} 
\label{alg:terraform}
\begin{algorithmic}[1]
\State \textbf{Input:} FL algorithm $\mathcal{A}$; Rounds $R$; Max. iterations $T$; Client set $\mathcal{C}$; Client dataset size $|D_{\text{train}}^k|$; No. of clients per round $K$; Min. clients required for splitting $\eta$
\State \textbf{Output:} Final global model $\theta_{R,0}$
\State \textbf{Initialize:} Global model $\theta_{0,0}$

\For{each round $r = 0, \dots, R-1$}
    \State \textbf{Initialize:} Randomly selected clients $\mathcal{C}_{r}^{K}$; initial global model $\theta_{r,0}$;
    \State $\mathcal{C}_{r,0}^{\mathcal{H}} \leftarrow \mathcal{C}_{r}^{K}$, where $\mathcal{H}:$ hard clients 
    \For{each splitting iteration $t = 0, \dots, T-1$}
        \State $\theta_{r,{t+1}}, \Delta w_{r,t}^{\mathcal{H}} \leftarrow \mathcal{A}(\theta_{r,t}, \mathcal{C}_{r,t}^{\mathcal{H}})$
        \State $\mathcal{C}_{r,t}^{\mathcal{H}(\text{sorted})}, \Delta w^{\mathcal{H}(\text{sorted})}_{r,t} \leftarrow \text{Sort}(\mathcal{C}_{r,t}^{\mathcal{H}}, \Delta w^{\mathcal{H}}_{r,t})$ \Comment{Sort clients}
        \State $k_{Q1},k_{Q3} \leftarrow \text{IQR}(D^{\mathcal{H}(\text{sorted})}_{\text{train}})$
        \State $\tau_{\text{split}} = \underset{k_{Q1} \le \tau < k_{Q3}}{\operatorname{argmin}} \left( \operatorname{Var}_{\text{intra}}(U_1,U_2) \right)$
        \State $\mathcal{C}_{r,t+1}^{\mathcal{H}} \leftarrow \mathcal{C}_{r,t}^{\mathcal{H}(\text{sorted})}[\tau_{\text{split}} :]$ 
        \If{$|\mathcal{C}_{r,t+1}^{\mathcal{H}}| < \eta$}
            \State \textbf{break}
        \EndIf
    \EndFor
    \State $\theta_{r+1,0} \leftarrow \theta_{r,t+1}$
\EndFor
\State \textbf{return} $\theta_{R,0}$
\end{algorithmic}
\end{algorithm}

\subsection{Terraform Algorithm}
We discuss the pseudo-algorithmic logic for Terraform in Algorithm~\ref{alg:terraform}. 
It takes as input a base FL algorithm $\mathcal{A}$, total rounds $R$, maximum splitting iterations $T$, the total client pool $\mathcal{C}$, client dataset sizes $|D_{train}^{k}|$, number of clients per round $K$, and minimum client threshold $\eta$.
The server initializes a global model $\theta_{0,0}$ (Steps 1-3). 

Each round $r$ starts by selecting a pool of K clients, which represents the initial set of 'hard' clients $\mathcal{C}^{\mathcal{H}}_{r,0}$ (Steps 5-6). 
For each iteration $t$, the algorithm $\mathcal{A}$ utilizes the current hard set $\mathcal{C}^{\mathcal{H}}_{r,t}$ and the global model $\theta_{r,t}$.
It returns the set of the clients' gradient updates, $\Delta w_{r,t}^{\mathcal{H}}$, and updated global model, $\theta_{r,t+1}$ (Step 7). 
This set of clients $\mathcal{C}^{\mathcal{H}}_{r,t}$ and $\Delta w_{r,t}^{\mathcal{H}}$, are then sorted according to the gradient update magnitudes, $|\Delta w^{\mathcal{H}}_{r,t}|$ (Step 8). 
To ensure a robust split, we find the first and third quartiles ($k_{Q1}, k_{Q3}$) indices using the sorted set of client data sizes, $D^{\mathcal{H\text{(sorted)}}}_{\text{train}}$ (Step 9).
The IQR first calculates the running sum for $D^{\mathcal{H\text{(sorted)}}}_{\text{train}}$ and then calculates $k_{Q1}$ and $k_{Q3}$ as described in~\ref{sec:iqr}.
The optimal split index $\tau_{split}$ is computed within the IQR indices (Step 10). 
The client set for the next iteration $\mathcal{C}^{\mathcal{H}}_{r,t+1}$ is then populated with clients ranging from the split index $\tau_{split}$ to the end of the set, $\mathcal{C}_{r,t}^{\mathcal{H(\text{sorted})}}$ (Step 11). 
The iterations end when client set size $< \eta$ (Steps 12-14).

The final model after $T$ iterations is used to initialize the following round's model $\theta_{r+1,0}$ (Step 16). 
The output at the end of $R$ rounds is the resulting global model $\theta_{R,0}$.

%% file: experiments.tex
\section{Evaluation}
\label{s:eval}

\subsubsection{Baselines.}
We evaluate Terraform against {\em five} state-of-the-art client selection methodologies: 
(i) Random Sampling i.e., FedAvg, which was presented in FedAvg; 
(ii) {\em PoC}; 
(iii) {\em Oort}; 
(iv) {\em HiCS-FL}; and 
(v) Baseline methodology in HiCS-FL ({\em HBase}), i.e., which was presented in FedProx. 

\subsubsection{FL Algorithms.}
We follow prior works on client selection methodologies and test all selection methodologies on two popular FL algorithms ($\mathcal{A}$), FedAvg and FedProx. 

\subsubsection{Datasets.}
We use the following {\em four} standard image classification datasets for benchmarking: 
(i) {\em CIFAR-10}: colored image dataset of objects, $10$ classes~\cite{cifar}; 
(ii) {\em CIFAR-100}: fine-grained version of CIFAR-10, $100$ classes~\cite{cifar}; 
(iii) {\em Tiny ImageNet}: colored image dataset of natural objects, $200$ classes~\cite{tinyimagenet}; and
(iv) {\em FEMNIST}: grayscale image dataset of handwritten characters, $62$ classes~\cite{femnist}. 

\subsubsection{Client Models.}
For CIFAR-10, each client used a 5L CNN model with $3$ convolutional layers with $3\times3$ kernels ($32/64/64$ output channels), followed by two $2\times2$-maxpooling layers (stride $2$), a fully connected layer with $64$ units, and 1D classification layer. 

For CIFAR-100, each client uses a 5L CNN architecture introduced by~\citet{flute} with $2$ convolutional layers with $5\times5$ kernel ($64/128$ channels), followed by $2\times2$-maxpooling (stride $2$) layers, followed by $3$ fully connected layers ($3200/256/128$ units with ReLU) and a 
softmax classification layer. 
For Tiny ImageNet, we used the pre-trained ResNet18 model~\cite{resnet}. 

For FEMNIST, we follow the architecture of FedAvg, i.e., the 4L CNN model with $2$ convolutional layers with $5\times5$ kernel ($32/64$ channels), each with a sequential block of $2\times2$-maxpooling (stride 2), followed by a $512$ unit fully connected layer (ReLU), and a softmax classification layer.

\begin{table*}[t]
\centering
\caption{Accuracy of client models with \textbf{best} and \underline{second-best} methodologies highlighted.}
\label{tab:performance_datasets}
\resizebox{0.7\textwidth}{!}{%
\begin{tabular}{|c|l|c|c|c|c|c|c|}
\hline
\multirow{2}{*}{$\mathcal{A}$}  & \multirow{2}{*}{Methodology}   & \multicolumn{3}{c|}{CIFAR-10} & CIFAR-100 & Tiny ImageNet & FEMNIST \\ \cline{3-5}
                                &                           & {1} & {2} & {3} & & & \\ \hline
\multirow{6}{*}{FedAvg}         & Random                    & \underline{67.76\%} & 58.21\% & 54.69\% & 15.26\% & 35.44\% & 81.28\%\\
                                & HBase                     & 54.76\% & 41.09\% & 41.69\% & 15.58\% & 35.99\% & 79.44\% \\
                                & PoC                       & 47.84\% & 27.37\% & 24.51\% & 13.73\% & 34.29\% & 80.31\% \\
                                & HiCS-FL                   & 65.57\% & 57.24\% & \underline{57.00\%} & 16.38\% & \underline{36.09\%} & \underline{81.46\%} \\
                                & Oort                      & 65.41\% & \underline{60.20\%} & 54.71\% & \underline{16.43\%} & 35.05\% & 80.38\% \\
                                & \textbf{Terraform (Ours)}          & \textbf{68.23\%} & \textbf{63.23\%} & \textbf{57.10\%} & \textbf{24.22\%} & \textbf{42.34\%} & \textbf{83.37\%}\\
\hline
\multirow{6}{*}{FedProx}        & Random                    & \underline{66.17\%} & \underline{60.01\%} & 52.47\% & 15.97\% & \underline{35.46\%} & 81.28\%\\
                                & HBase                     & 55.90\% & 41.21\% & 43.79\% & 16.71\% & 35.10\% & \underline{81.49\%}\\
                                & PoC                       & 47.33\% & 20.28\% & 28.00\% & 13.81\% & 34.99\% & 80.31\%\\
                                & HiCS-FL                   & 67.10\% & 58.59\% & \underline{57.94\%} & \underline{17.54\%} & 35.36\% & 81.08\%\\
                                & Oort                      & 65.83\% & 58.58\% & 54.58\% & \underline{17.54\%} & 35.31\% & 80.59\%\\
                                & \textbf{Terraform (Ours)}          & \textbf{68.52\%} & \textbf{62.73\%} & \textbf{59.13\%} & \textbf{24.81\%} & \textbf{42.44\%} & \textbf{83.52\%}\\
\hline
\end{tabular}%
}
\end{table*}

\begin{table*}[t]
\centering
\caption{Accuracy of client models for FMNIST scenarios with \textbf{best} and \underline{second-best} methodologies highlighted.}
\label{tab:performance_fmnist}
\resizebox{0.78\textwidth}{!}{%
\begin{tabular}{|c|l|c|c|c|c|c|c|c|c|}
\hline
$\mathcal{A}$               & Methodology                        & 1 & 2 & 3 & 1* & 2* & 3* & 4 & 5\\
\hline
\multirow{6}{*}{FedAvg}     & Random                        & 85.62\% & \underline{83.28}\% & \underline{70.19}\% & \underline{84.05}\% & 83.84\% & 70.15\% & 85.18\% & 84.78\%\\
                            & HBase                         & 74.23\% & 71.20\% & \textbf{71.69\%} & 80.74\% & 83.11\% & \textbf{73.15\%} & 85.36\% & 85.70 \% \\
                            & PoC                           & 80.29\% & 79.36\% & 58.11\% & 81.93\% & 83.17\% & 53.40\% & 85.32\% & 84.93\% \\
                            & HiCS-FL                       & \underline{87.14}\% & 82.81\% & 68.14\% & 83.60\% & 84.32\% & 71.18\% & \underline{85.39}\% & 85.16\% \\
                            & Oort                          & \textbf{89.01\%} & \textbf{85.42\%} & 59.52\% & 83.21\% & \underline{84.75}\% & 57.39\% & 85.09\% & \underline{85.79}\% \\
                            & {\bf Terraform (Ours)}        & 85.62\% & \underline{83.28}\% & \underline{70.19}\% & \textbf{85.49\%} & \textbf{85.47\%} & \underline{71.21}\% & \textbf{86.08\%} & \textbf{86.51\%} \\    \hline
\multirow{6}{*}{FedProx}    & Random                        & \underline{85.78}\% & \underline{83.24}\% & \underline{70.97}\% & 83.27\% & 83.99\% & 70.01\% & 85.22\% & 85.10\% \\
                            & HBase                         & 76.23\% & 72.39\% & \textbf{72.11\%} & 81.35\% & 83.07\% & \textbf{71.37\%} & 85.75\% & \underline{85.70}\% \\
                            & PoC                           & 84.50\% & 79.67\% & 59.42\% & 82.22\% & 83.08\% & 65.42\% & 85.44\% & 84.99\% \\
                            & HiCS-FL                       & 86.00\% & 80.61\% & 69.48\% & 83.41\% & \underline{85.10}\% & 69.71\% & \underline{85.77}\% & 84.84\% \\
                            & Oort                          & \textbf{87.11\%} & \textbf{85.43\%} & 59.42\% & \underline{83.47\%} & 84.69\% & 66.79\% & 85.47\% & 85.51\% \\
                            & {\bf Terraform (Ours)}        & \underline{85.78}\% & \underline{83.24}\% & \underline{70.97}\% & \textbf{83.91\%} & \textbf{85.75\%} & \underline{70.02}\% & \textbf{87.24\%} & \textbf{86.63\%} \\
\hline
\end{tabular}%
}
\end{table*}

\subsubsection{Dataset Partitioning.}
To simulate a realistic federated setting with statistical heterogeneity, we follow the data partitioning schemes of HiCS-FL. 
These schemes generate each client’s local dataset using a Dirichlet distribution~\cite{dirichlet}.
For example, consider $100$ clients and a predefined set of five $\alpha$ values.
Clients are divided into $20$ subsets ($100/5 = 20$), with each subset chronologically assigned one of the $\alpha$ values. 
Each client in a subset receives its local data allocation through the Dirichlet distribution parameterized by its assigned $\alpha$. 
A smaller $\alpha$ produces a higher class imbalance, resulting in greater heterogeneity among clients’ data. 
Thus, given a total dataset of $1000$ samples, the Dirichlet distribution assigns samples to clients according to their designated $\alpha$, simulating varying levels of data heterogeneity across the client set.

We use HiCS-FL partitioning scenarios. 

FMNIST uses $50 \text{-} 100$ clients, and its following $8$ scenarios are based on $5$ $\alpha$ values and $3$ sampling rates.
Scenarios for sampling rate = 0.1, $50$ clients,
\textbf{(1)}: $\alpha$ = $\{$0.001, 0.002, 0.005, 0.01, 0.5$\}$, 
\textbf{(2)}: $\alpha$ = $\{0.001, 0.002, 0.005, 0.01, 0.2\}$, and
\textbf{(3)}: $\alpha$ = $\{0.001\}$.
Scenarios for sampling rate = 0.3, $50$ clients,
\textbf{(1\textsuperscript{*})}: $\alpha$ = $\{$0.001, 0.002, 0.005, 0.01, 0.5$\}$, 
\textbf{(2\textsuperscript{*})}: $\alpha$ = $\{$0.001, 0.002, 0.005, 0.01, 0.2$\}$, and
\textbf{(3\textsuperscript{*})}: $\alpha$ = $\{$0.001$\}$, 
Scenarios for sampling rate = 0.15, $100$ clients,
\textbf{(4)}: $\alpha$ = $\{$0.1, 0.1, 0.1, 0.3, 0.3$\}$, and
\textbf{(5)}: $\alpha$ = $\{$0.05, 0.1, 0.15, 0.2, 0.25, 0.3, 0.35, 0.4, 0.45, 0.5$\}$.

CIFAR-10 uses $50$ clients, and its following $3$ scenarios are based on sampling rate = 0.2 and $3$ $\alpha$ values.
\textbf{(1)}: $\alpha$ = $\{$0.001, 0.01, 0.1, 0.5, 1$\}$, 
\textbf{(2)}: $\alpha$ = $\{$0.001, 0.002, 0.005, 0.01, 0.5$\}$, and 
\textbf{(3)}: $\alpha$ = $\{$0.001, 0.002, 0.005, 0.01, 0.1$\}$. 

As HiCS-FL does not evaluate on datasets CIFAR-100, Tiny ImageNet, and FEMNIST, 
we use the following scenarios:
CIFAR-100 uses $250$ clients and sampling rate = 0.128, whereas Tiny ImageNet uses $500$ clients and sampling rate = 0.058; $\alpha$ = 0.1 for both.
FEMNIST has a total of $800$k+ samples, and these samples were assigned among $3597$ clients using the writer ID~\cite{fedprox} partitioning. We used a sampling rate of $0.011$.

\subsubsection{Setup.}
We use NVIDIA's A100 Tensor Core GPUs to run our experiments. 
Each result is averaged over three runs to eliminate any noise.

We follow HiCS-FL and FedProx and set the hyper-parameter $\mu$ of the regularization term of FedProx to 0.1.
For FMNIST dataset, we use the stochastic gradient descent (SGD) optimizer. 
The learning rate is initially set to $0.001$ and decreases every $10$ iterations with a decay factor of $0.5$. 
The global communication rounds is set to $200$. 
The number of local epochs is set to $2$ and the size of a mini-batch is set to $64$. 
For datasets CIFAR-10, CIFAR-100, and Tiny ImageNet, we use the popular optimizer Adam~\cite{kingma2015adam} with a learning rate of $0.001$ ($0.01$ for Tiny ImageNet), a decay factor of $0.5$, global communication rounds of $100$ ($500$ for CIFAR-10), local epochs of $2$, and a mini-batch size of $64$. 
For FEMNIST, we use SGD optimizer with a learning rate of $0.01$, local epochs of $5$, and a mini-batch size of $32$.

\subsection{Results}
We now illustrate the results of comparing \tf{} against other client selection methodologies. 
We use Tables~\ref{tab:performance_datasets} and~\ref{tab:performance_fmnist} to illustrate the average accuracy of the global model on the test data of the participating clients using FL algorithms.


In Table~\ref{tab:performance_datasets}, we observe that \tf{} yields the highest accuracy among all the methodologies for all cases.
For CIFAR-10, as expected, the accuracy for all methodologies decreases with increasing heterogeneity. 
Consequently, on FedAvg, we observe that Oort and HiCS-FL yield better accuracy than Random for scenarios 2 and 3, and HiCS-FL does better than Oort on scenario 3 due to its heterogeneity-focused client clustering.
{\em Note:} for CIFAR-10 and FMNIST, we follow the three scenarios of HiCS-FL (1, 2, and 3), which restrict us to initially sample only $10$ and $5$ clients, respectively.
If Terraform is allowed to sample more initial clients in each round, it can yield higher accuracy, which is evident in Table~\ref{tab:performance_fmnist} (4 and 5).
However, on FedProx, Random outperforms others on scenario $2$ because FedProx's high regularization factor ($\mu = 0.1$) leads to underfitting.

On Tiny ImageNet and CIFAR-100, \tf{} shines the most; it achieves up to $ 17.31\%$ and $47.41\%$ more accuracy than the second-best methodology as it allowed to sample a higher number of clients initially in each round.
%
Although a similar trend for \tf{} exists on FEMNIST, we attribute it to more classes in the dataset, which increases non-iid relationships, and increase the data heterogeneity.

In Table~\ref{tab:performance_fmnist}, we evaluate \tf{} on $8$ FMNIST scenarios.
As scenarios 1, 2, and 3 are ported from HiCS-FL, we force \tf{} to sample only $5$ clients initially.
Consequently, \tf{} is able to run only one iteration per round, which is same as running Random; thus, similar accuracies. 
For rest of the scenarios, \tf{} does not have this restriction, and thus, it continues outperforming other methodologies.
One exception, however, is scenario 3*, where \tf{} is the second-best because sampling even $15$ clients initially, in each round, is not sufficient; 
we hypothesize that \tf{} needs to sample a much higher number of clients. 

\begin{figure}[t]
  \begin{minipage}{0.49\columnwidth}
    \centering
    \captionsetup{skip=0pt} 
    \caption*{(a) CIFAR100}  
    \includegraphics[width=\textwidth]{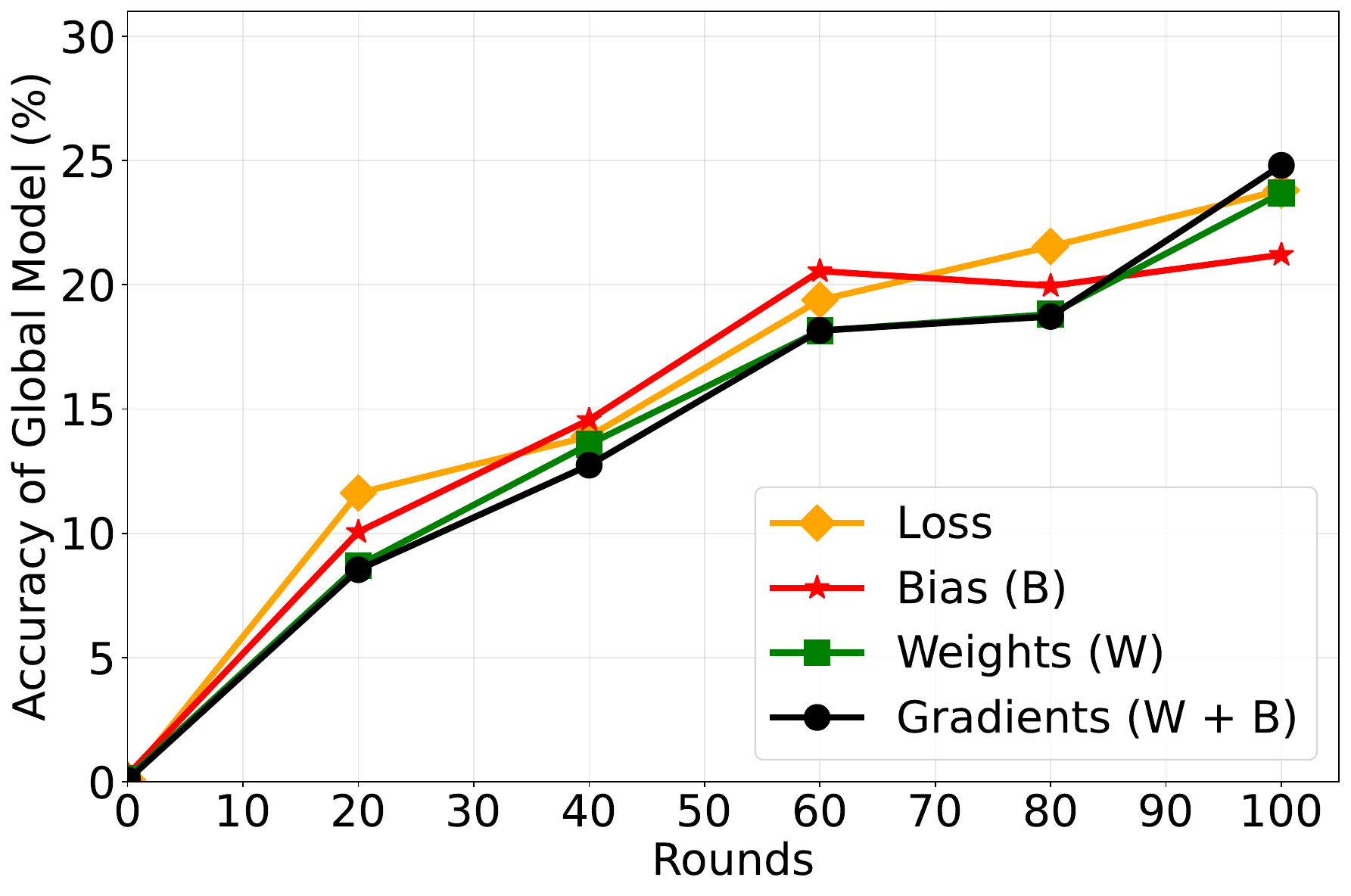}
  \end{minipage}
  \begin{minipage}{0.49\columnwidth}
    \centering
    \captionsetup{skip=0pt}
    \caption*{(b) Tiny ImageNet}
    \includegraphics[width=\textwidth]{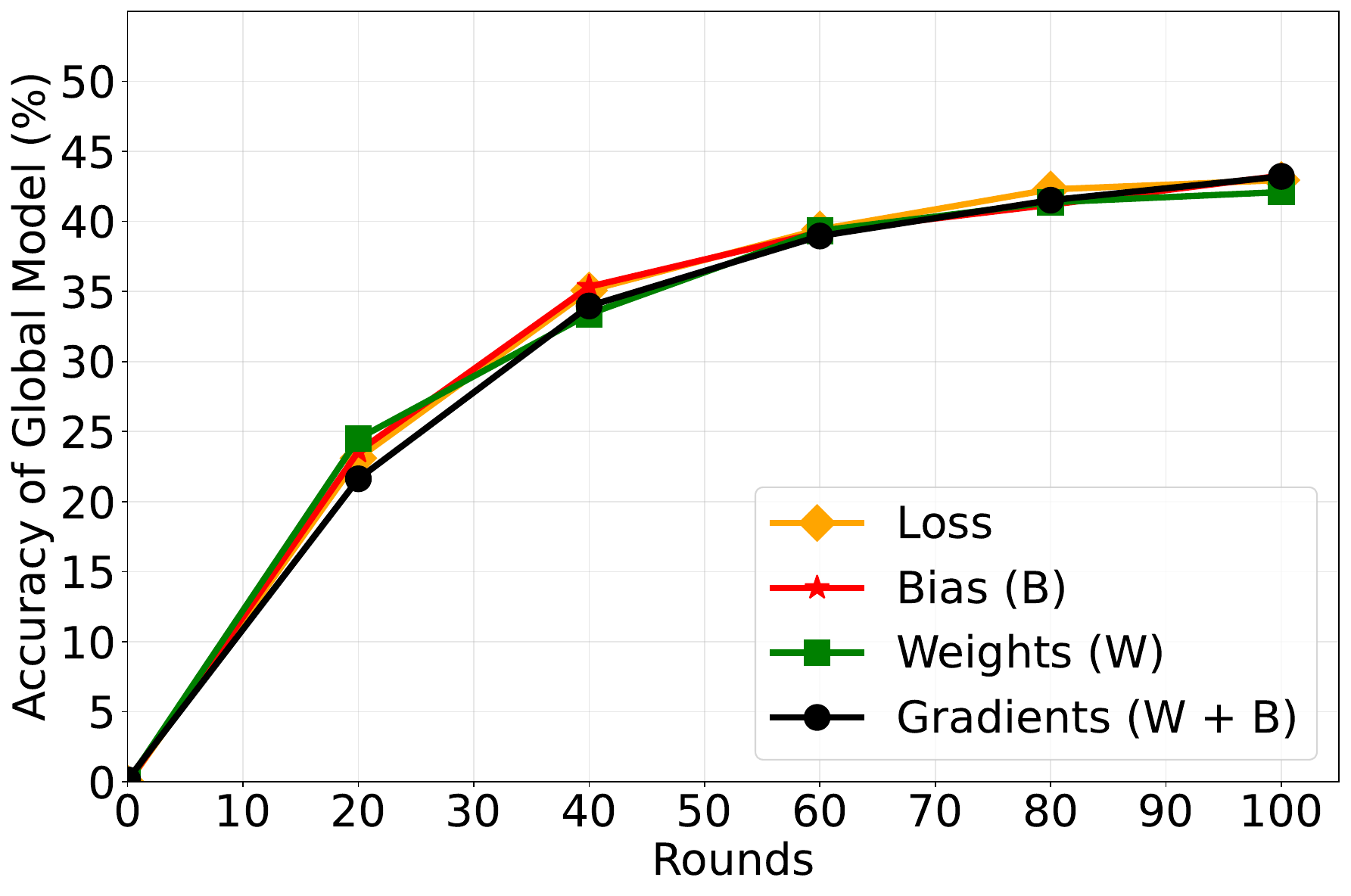}
  \end{minipage}
  \captionsetup{skip=0pt}
  \caption{Accuracy of model on varying client updates.}
  \label{fig:gradient-ablation}
\end{figure}

\subsection{Ablation Studies}
We validate Terraform's contribution through ablation studies that assess the efficacy of client updates, IQR, and client threshold ($\eta$) in client selection.

\subsubsection{Varying Client Updates.} 
\tf’s gradient updates are, in essence, a combination of the weights and biases from the clients’ models’ final layers. 
In Figure~\ref{fig:gradient-ablation}, we illustrate the accuracy of \tf{} when, instead of using gradient updates, it uses (i) loss, (ii) biases, and (iii) weights.

We observe that using gradient updates (weights + bias) with \tf{} yields the best performance at the end of $100$ rounds of training for both CIFAR-100 and Tiny ImageNet.
Moreover, using biases, which is a HiCS-FL approach, has the lowest performance for the CIFAR-100 dataset compared to the Tiny ImageNet due to the reduced number of classes in the former dataset.
This, in turn, reduces the bias values per client, thereby affecting the representative values for the client's distribution.



\begin{figure}[t]
  \begin{minipage}{0.49\columnwidth}
    \centering
    \captionsetup{skip=0pt}
    \caption*{(a) CIFAR100}
    \includegraphics[width=\textwidth]{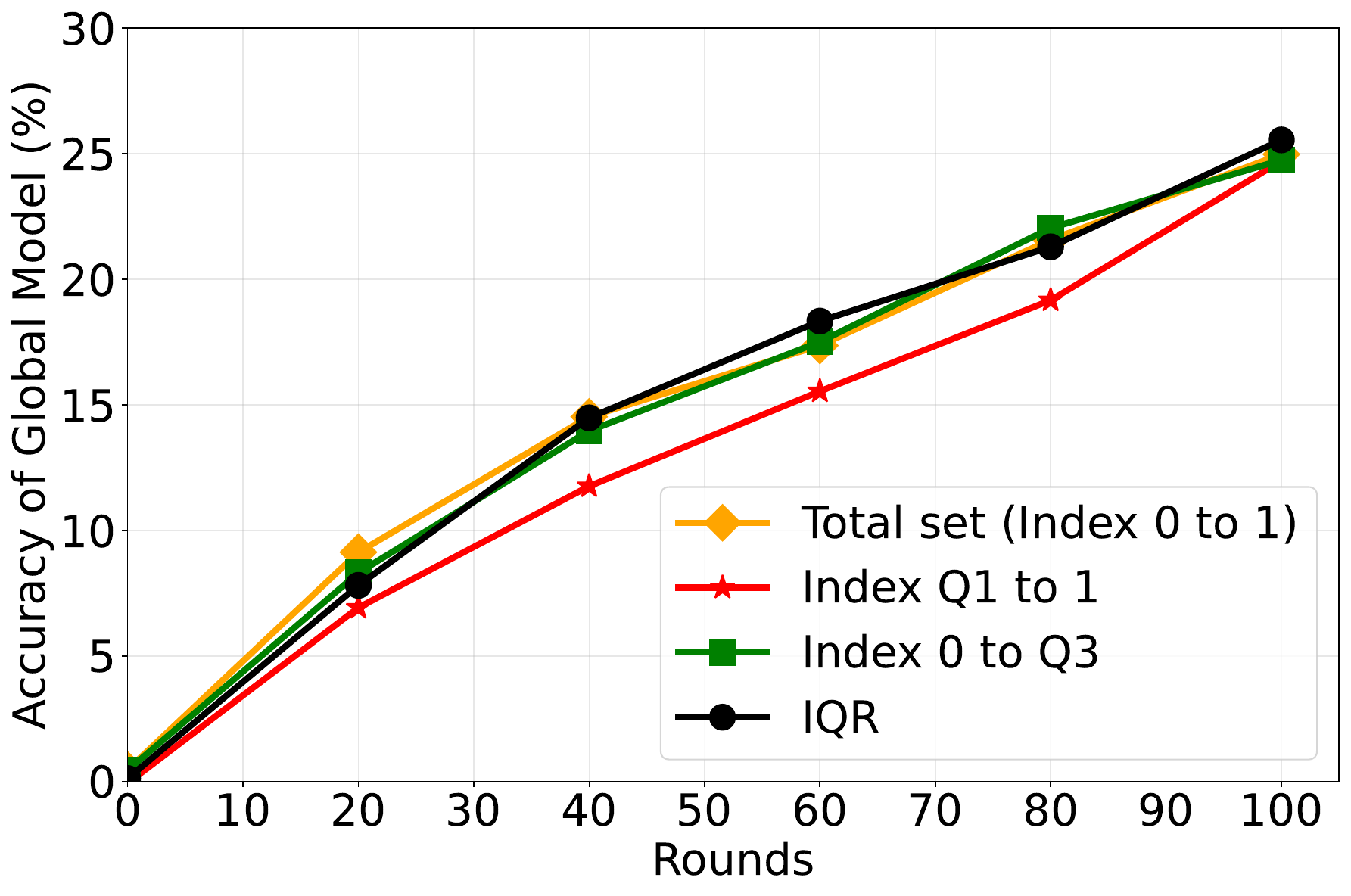}
  \end{minipage}
  \begin{minipage}{0.49\columnwidth}
    \centering
    \captionsetup{skip=0pt}
    \caption*{(b) Tiny ImageNet}
    \includegraphics[width=\textwidth]{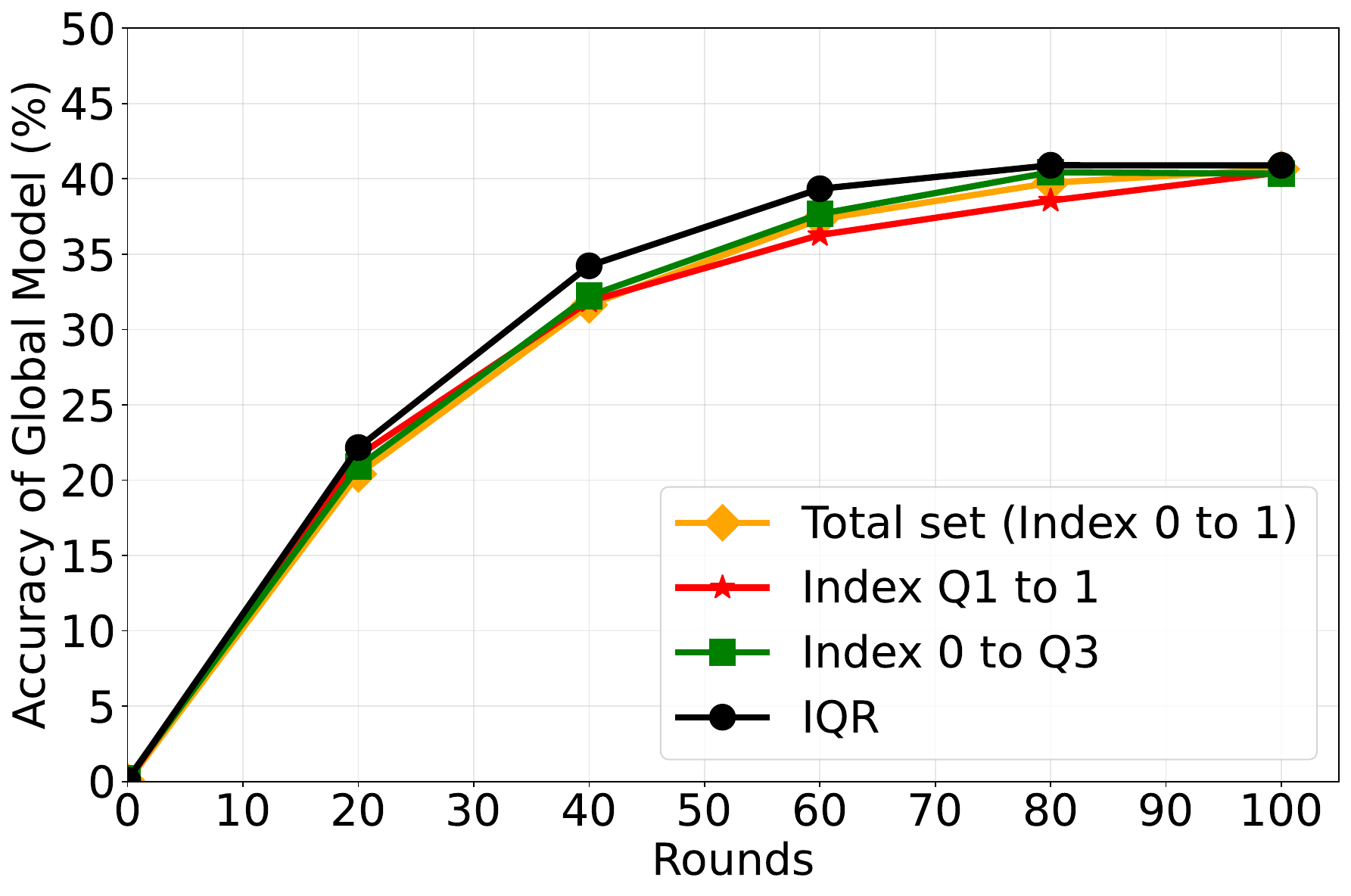}
  \end{minipage}
  \captionsetup{skip=0pt}
  \caption{Accuracy of global model on varying quartiles.}
  \label{fig:iqr-ablation}
\end{figure}

\begin{figure}[t]
  \centering
  \includegraphics[width=0.8\columnwidth]{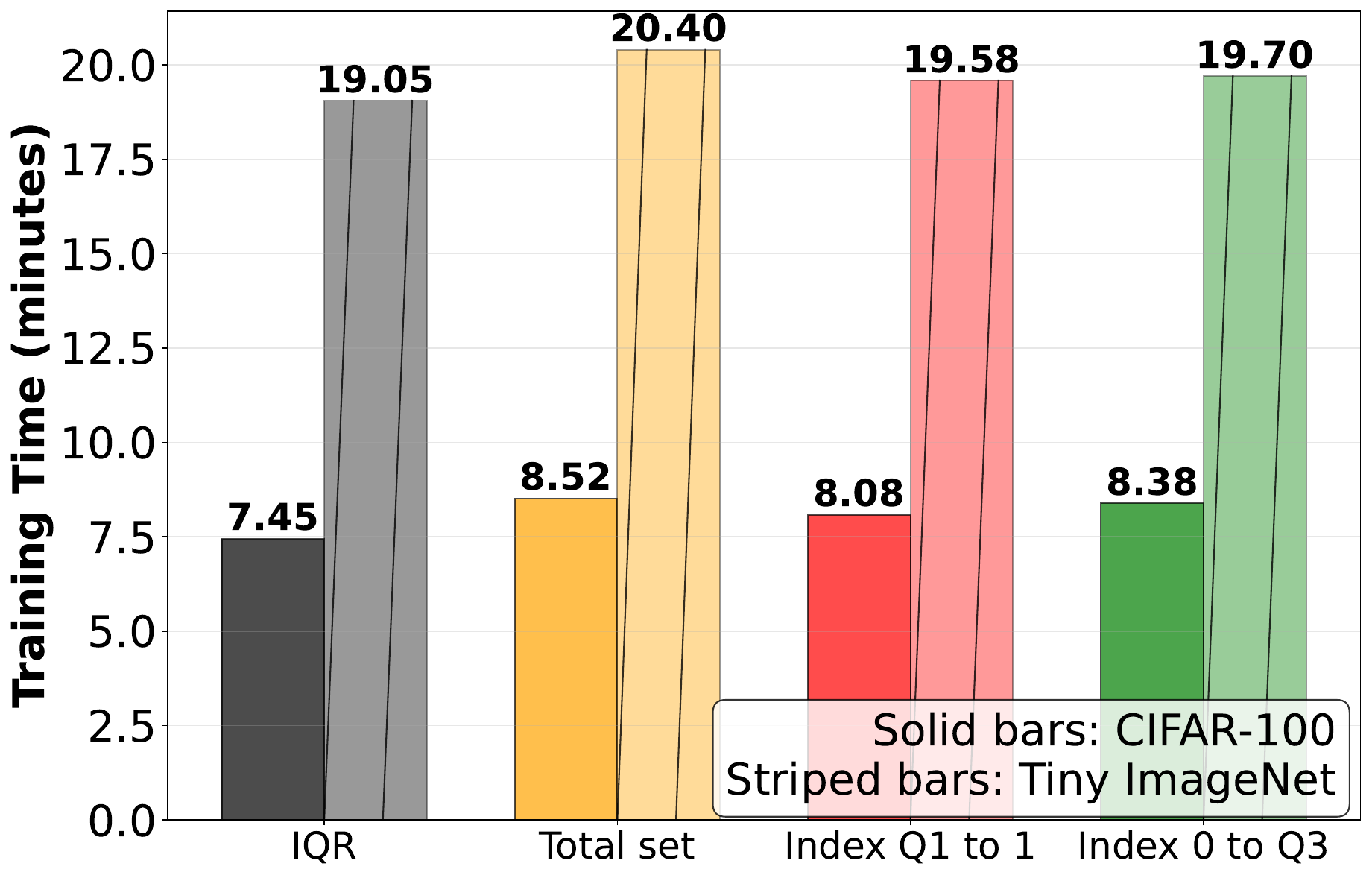}
  \captionsetup{skip=0pt}
  \caption{Training time on varying quartiles.}
  \label{fig:compute-ablation}
\end{figure}

\begin{table}[t]
\centering
\caption{Accuracy on varying thresholds, $\eta$.}
\label{tab:threshold-ablation}
\resizebox{0.7\columnwidth}{!}{%
\begin{tabular}{|l|c|c|c|}
\hline
Dataset         & $\eta=2$ & $\eta=3$ & $\eta=4$   \\ \hline
FEMNIST         & 83.61\% & 83.15\% & 83.52\%             \\ \hline
FMNIST(1*)      & 80.26\% & 82.58\% & 83.91\%             \\ \hline
FMNIST(4)       & 86.07\% & 86.51\% & 87.24\%             \\ \hline
\end{tabular}
}
\end{table}

\subsubsection{Varying Quartiles.} 

To determine the optimal split index $\tau_{split}$, Terraform searches in its Inter-Quartile Range (IQR) for an index that minimizes the intra-split variance ($\operatorname{Var_{intra}}$).
A brute-force approach to determine the split index $\tau_{split}$ is to iterate over each index in the set of client dataset sizes, $D^{\mathcal{H\text{(sorted)}}}_{\text{train}}$, and calculate the intra-split variance. 
Such an approach would be computationally expensive, and IQR minimizes this cost by reducing the search space: 
from the total set (0,1) to the region between the first and third quartiles of the set (Q1, Q3).

In Figure~\ref{fig:iqr-ablation}, on the CIFAR100 and Tiny ImageNet datasets for $100$ rounds, we compare IQR (Q1, Q3) against three other splitting ranges:
(0, 1), (0, Q3), and (Q1, 1). 
Our results illustrate that IQR is not only sufficient for determining split index $\tau_{split}$, but also helps to discard outliers that reduce the test accuracy. 
%
Additionally, (Q3, 1) have the poorest performance, which validates the theory that hyper-focusing on only highly heterogeneous clients overfits the model and reduces generalization.

{\bf Training time:} In Figure~\ref{fig:compute-ablation}, we measure the training time complexity for the 4 quartile ranges and observe that our claims that IQR and total set have the lowest and highest training time, respectively, hold.

\subsubsection{Threshold for $\#$Clients ($\eta$).}
\tf's hierarchical sampling approach iteratively splits the set of clients into easy $\mathcal{E}$ and hard $\mathcal{H}$ clusters. 
Specifically, we keep splitting the hard cluster until the following termination condition is met: the size of the hard cluster falls below a predefined threshold ($\eta$).
Determining an optimal value of $\eta$ saves compute as further splitting will yield minimal benefits.
Thus, we set $\eta$ to $4$.
%
In Table~\ref{tab:threshold-ablation}, we illustrate the accuracy on three different values for $\eta=\{2, 3, 4\}$ on FEMNIST and FMNIST (1 and 4*) scenarios. 
We observe that $\eta = 4$ performs well on both datasets.
Although $\eta = 2$ yields a marginally better accuracy on the FEMNIST dataset, $\eta = 4$ reduces the computational cost for training and yields only $0.11\%$ lower accuracy than $\eta = 2$, which makes it an optimal choice.

%% file: concl.tex
\section{Conclusions}
In this paper, we presented \tf, a novel client selection methodology that mitigates the impact of heterogeneous clients on federated learning. 
Terraform achieves this goal by hierarchically selecting and retraining heterogeneous clients. 
Terraform performs this hierarchical selection on the basis of clients' local model's final-layer gradient updates, which it uses to determine the optimal 
inter-quartile range (IQR) to split clients into easy and hard clusters.
Terraform retrains clients in hard clusters and our results illustrate that it outperforms existing works by up to $47\%$.